\documentclass[manuscript]{aastex}

\newcommand{\VRI}{\hbox{$V\!RI$}}
\newcommand{\VR}{\hbox{$V\!-\!R$}}

\newcommand{\RI}{\hbox{$R\!-\!I$}}

\newcommand{\kms}{km~s$^{-1}$}

\newcommand{\hii}{\ion{H}{2}}
\newcommand{\msun}{$M_\odot$}

\shortauthors{Dolan \& Mathieu}
\shorttitle{Photometric Study of the $\lambda$ Orionis Region}

\begin{document}

\title{A Photometric Study of the Young Stellar Population
Throughout the $\lambda$ Orionis Star-Forming Region}

\author{Christopher J. Dolan\altaffilmark{1,2} \and 
  Robert D. Mathieu\altaffilmark{1}}
\altaffiltext{1}{Department of Astronomy, University of
  Wisconsin--Madison, 475 North Charter Street, Madison, WI 53706}
\altaffiltext{2}{Visiting Astronomer, Kitt Peak National Observatory,
  National Optical Astronomy Observatories, which is operated by the
  Association of Universities for Research in Astronomy, Inc.  (AURA)
  under cooperative agreement with the National Science Foundation.}
\email{dolan@astro.wisc.edu; mathieu@astro.wisc.edu}

\begin{abstract}

We present {\VRI} photometry of 320,917 stars with $11 \lesssim R
\lesssim 18$ throughout the $\lambda$~Ori star-forming region.  Using
the more spatially limited spectroscopic surveys of Dolan \& Mathieu to
define the color - magnitude domain of young low-mass members of the
association, and removing statistically 
the field stars in this domain, we use our
photometry to identify a representative PMS population 
throughout the interior of
the molecular ring.  The spatial distribution
of this population shows a concentration of PMS stars around
$\lambda$~Ori and in front of the B35 dark cloud.  However, 
few PMS stars are found outside
these pockets of high stellar density, 
suggesting that star formation was concentrated in an elongated
cloud extending from B35 through $\lambda$~Ori to the B30 cloud.

We find a lower limit for the global stellar mass of about 500 {\msun}.
We find that the global ratio of low- to high-mass stars is similar to
that predicted by the field initial mass function, but this ratio varies strongly as a
function of position in the star-forming region. Locally, the star-formation
process does not produce a universal initial mass function.

Using our derived stellar ages across the region, we
construct a history of the star-forming complex.  This history
incorporates a recent supernova to explain the distribution of stars and
gas today.  We infer that most of the present molecular ring was formed
by ejecta from the center driven by the supernova blast about 1 Myr ago.
However, we suggest that the B30 and B35 clouds were primordial, and 
massive enough to be mostly
little disturbed by the shock.  The stars which we see today trace the
former extent of the cloud complex.  Given the kinematics of the
stellar population, we predict that the association will disperse into
the field within a few tens of Myr.  The gas will be dispersed on a
similar time scale, or faster if $\lambda$~Ori becomes a supernova
before it escapes the region.

\end{abstract}


\section{Introduction}

Much of our understanding of the process of low-mass star formation
arises from studies of individual young stars.  Historically, the
most popular regions for the study of these stars are nearby dark
clouds, such as
Taurus-Auriga, Chameleon, Lupus and $\rho$ Oph. In these regions,
the T Tauri stars interact with the local interstellar medium 
with parsec-long jets and
substantial molecular outflows, but their spheres of influence are
usually smaller than their separations.  Thus models of star
formation which arise from observations of stars in such T
associations often consider only stars in isolation.

Recent advances in technology have permitted detailed studies of
low-mass stellar populations in OB associations.
The most massive members of such OB
associations have large spheres of influence and may assail the entire
star-forming region with dissociating or ionizing radiation or
disrupting winds.  Thus the massive stars must be integrated into the
formation process of associated low-mass stars.

Can the massive stars terminate nearby star formation by destroying
the gas clouds?  Alternatively, can they enhance star formation by
compressing the clouds?  Could both of these effects occur in
different regions of the same star-forming complex?  Can these effects
change the resulting spatial and mass distributions of young stars?

Recent studies are beginning to answer affirmatively to all of these
questions.  In cases where the massive stars ionize clouds and reveal
embedded stars (e.g.\ Eagle Nebula, \citealt{hes96}) or evaporate
circumstellar disks of existing stars (e.g.\ Trapezium,
\citealt{bal98}), it is clear that massive stars have {\em terminated}
star formation.  On the other hand, \citet{wal94} and \citet{pre99}
show that star formation in the Upper Scorpius OB association was
likely {\em triggered} by a nearby supernova, since all the stars were
formed within a very short timespan.

For many years, star formation was hypothesized to be bimodal:
low-mass and high-mass stars were thought to form under very different
conditions (c.f.\ \citealt{shu88}).  With the notable exception of the
Trapezium, evidence for low-mass stars near OB stars was lacking until
the 1980s.  Since then, however, young low-mass stars have been shown
to be ubiquitous.  In particular the ROSAT All-Sky Survey found young,
low-mass stars associated with OB stars throughout the Gould Belt
(\citealt{gui98}).  Since this realization, surveys have demonstrated
that the initial mass functions (IMFs) 
of several OB associations are consistent with the field
IMF (e.g., \citealt{pre99} in Upper Sco, \citealt{hil97} in the
Trapezium).  But these statements are for the global IMF of the star
formation regions.  In many associations, the OB stars are
more spatially concentrated than the rest of the stars.  Except perhaps in very
high density environments like the Trapezium, these associations are
too young for dynamical mass segregation.  Perhaps
some aspect of the star-forming process skews the local IMF towards
high-mass stars in some locations and low-mass stars in others.

To make a comprehensive exploration of the effect of massive stars on
low-mass star formation, this paper employs a photometric survey for 
young low-mass stars
throughout the $\lambda$~Orionis OB association.  The $\lambda$~Ori
region is a superb laboratory for studying the evolution of a
molecular cloud into an association of low- and high-mass stars.  The
star-forming complex contains a tight knot of OB stars encircled by a
40 pc diameter ring of dense molecular gas and dust \citep[see
Figure~15 of][]{dol00}.  Notable features of this clumpy ring include
the large B30 cloud on the northwest edge and the elongated B35 cloud
protruding inward from the eastern side.  By contrast, the interior of
this ring is nearly devoid of dense gas.  It is this relative
transparency that has made the $\lambda$~Ori region both understudied
in the past and appealing for the present study of the young stellar
population: the lack of opacity reveals the entire member population
as well as vast numbers of background field stars.  Until recently,
comprehensive studies have been heavily biased towards stars
with strong emission features such as H$\alpha$ \citep{due82} or
X-rays \citep[and several subsequent works]{ste95}.

In \citet[hereafter Paper~I]{dol99} and \citet[hereafter
Paper~II]{dol00}, we studied a spatial subset of the $\lambda$~Ori
region, focusing on a 6{\arcdeg}-long region extending 
from B30 to B35 through the
central concentration of OB stars.  For nearly every star in that
region with $12<R<16$ and redder than a 30 Myr \citet{dan94} isochrone
(3618 stars), we obtained a high-resolution spectrum with the
WIYN\footnote{The WIYN Observatory is a joint facility of the
University of Wisconsin-Madison, Indiana University, Yale University,
and the National Optical Astronomy Observatories.} Multi-Object
Spectrograph (MOS).  From these, we identified 266 pre-main-sequence
(PMS) stars via the
presence of strong lithium $\lambda$6708 absorption, a secure
diagnostic of youth.  Using model evolution tracks to determine ages,
we found that low-mass star formation started simultaneously
with the OB stars 6-8
Myr ago and has accelerated since then.  However, we found a marked
difference in recent star formation rates at different locations: 
within 2{\arcdeg} of the OB stars, there are
almost no stars younger than 1 Myr while further away stars of this
young age are plentiful.  These differing age distributions led to
our conclusion that there was a supernova 1 Myr ago which disrupted
further star birth in the vicinity of the OB stars.  
The model evolution tracks also yielded masses
for the PMS stars.  Compared with the massive OB stars, we concluded
that globally the IMF of the $\lambda$~Ori star-forming region was
consistent with the field IMF, but there were significant local
variations: the high-mass stars dominated the central region while
low-mass stars dominated at larger radii.  Those low-mass stars
closest to the OB concentration also show a remarkable lack of
H$\alpha$ emission at all ages.  We attributed this to FUV radiation
from the massive stars destroying the circumstellar disks of those
stars.

The age and mass data for the PMS stars showed us that nearly all of
the detected PMS stars lie in a region of the {\RI} versus $R$
color-magnitude diagram brighter and redder than the bulk of the field
stars.  This result suggested that photometric analysis would be an
effective tool for identifying and removing well over 99\% of the field
stars.  To exploit this feature, we have imaged most of the
$\lambda$~Ori region in {\VRI} and produced a comprehensive stellar
photometric database.  With these data, we identify all stars in the
region of the CMD where PMS members of the association are likely to
lie.  Using the subset of fields where we have spectroscopic data
(Paper 2), we construct a model of the field star distribution.
Statistically removing the field stars, we are left with a stellar
sample representative of the true PMS population.

This representative 
sample allows us to examine the spatial distribution
of the young population.  
This sample also permits us to count the total number of
low-mass members, and thereby compare the association IMF to the field
IMF.  In combination with the high- and low-mass
stellar age distribution derived in Paper~II, we develop in this paper
a comprehensive history of star formation in the $\lambda$~Ori
association.

\section{Data}

In \S~2 of Paper~II we presented a summary of our broad-band
photometric survey of the $\lambda$~Ori region.  This paper presents a
more detailed description and analysis of those photometric data.
Unlike Paper~II, we discuss only the ``1999 photometry'' in the
following.  In fact, some stars with older photometry from 1997 and
1998 are included in the dataset presented here where we have rare,
accidental holes in our 1999 sky coverage, but this accounts for just
0.1\% (361 of 320917) of the stars, so we ignore those older datasets
in the following discussion.

\subsection{Acquisition}

During December 15--20, 1999 we used the Mosaic imager on the KPNO
0.9m telescope to survey about 60 degrees$^2$ of the $\lambda$~Ori
star-forming region in {\VRI}.  This region includes the central
concentration of OB stars, portions of the encircling ring of
molecular clouds, and the nearly gas-free region in between.

The Mosaic imager is highly effective for such a survey with its
nearly 1{\arcdeg} field of view and pixel size of 0.42{\arcsec}.  In
total, we observed 61 fields with {\VRI} filters.  Table~\ref{tab:obs}
summarizes the conditions of the observing run.  A more detailed
analysis of individual fields from each night is included in
\S~\ref{sec:data-quality}.  In general, the run was of high quality,
but thin clouds affected a few fields, as described below.  The full
moon was very close to the star-forming region on the last nights,
causing the signal-to-noise for stars of a given magnitude to be
lower.


Prior to the run, we had selected 83 fields to observe, but
anticipating that weather would prevent us from finishing all of
these, we observed the fields in a priority order based on the
following criteria: fields which had been observed in 1997 and 1998
(to provide improved photometry for those crucial regions), fields
adjacent to and overlapping the 1997 and 1998 fields, fields which
overlapped archival ROSAT pointings (for future comparisons with the
X-ray data), and regions which appeared interesting in the
\citet{lan98} CO map (particularly, the B223 cloud and the small,
elongated cloud south of B35).



Figure~\ref{fig:nightmap} shows the location of the observed {\VRI}
fields.  The field numbers marked on Figure~\ref{fig:nightmap}
indicate our priority order in the sense that a lower number is higher
priority.  Table~\ref{tab:chxref} presents a cross-reference to the
field names used in Paper~II.  We present the center coordinates for
these fields as observed 
in Table~\ref{tab:fields}, as well as the night on which
the field was observed (cross-referenced to Table 1). (For those fields
not observed, a 0 is given for night observed and the intended field
center is provided for possible future extension of the survey.)  
Due to problems with
telescope pointing, the actual field centers are often 15--20{\arcsec}
(and up to 1.3{\arcmin} for some fields on Night 1) away from the
intended coordinates.  Since we usually overlapped adjacent fields by
5{\arcmin}, this is typically not a problem in terms of areal coverage.


For each observed field, we took three exposures per filter, each
shifted 100 pixels (about 43{\arcsec}) in Right Ascension and
Declination.  This is the minimum number needed to cover the inter-CCD
gaps in the Mosaic imager.  For almost 97\% of the image, where there
are no gaps or bad pixels, the multiple exposures increase the signal
by three when combined.  For just over 3\% of the field, a star will
lie on a gap in one of the three exposures.  Our photometry of such a
star would only use signal from the other two, which corresponds to a
19\% signal-to-noise degradation relative to the regions with full
coverage.  Finally, 0.1\% of the field lies on the intersection of
gaps from two images, which translates to a 43\% lower signal-to-noise
than when all three images can be used.

For each exposure, we used the following exposure times: 18 sec in
$I$, 15 sec in $R$, 27 sec in $V$.  These times were chosen to yield
signal-to-noise better than 100 in {\VRI} in the coadded images for a
star with $R=16$ and {\VR} = 1, while preventing $R=12$ stars from
saturating.

For photometric calibration, we took single {\VRI} exposures of 7--10
fields each night containing \citet{lan92} standard stars.  We
achieved nearly uniform sampling in airmass from at least 1.2 to 2.0
each night.  We always observed one or more of these calibration
fields before the first and after the last science field of each
night.

We also obtained ten bias exposures, five dome flats and a few
twilight flats per filter each night.  Over the course of the six
nights, we gathered enough twilight flats to make one high
signal-to-noise flat in each filter.  We discuss the relative merits
of dome and twilight flats in the next section.

\subsection{Reduction}
\label{sec:data-reduction}

We reduced all of our images with IRAF-MSCRED V3.2.3, closely
following the data reduction guide written for V2.0 by F.\ Valdes.
First, for each night we combined the bias images into a single image.
It is clear from these images that the bias voltages for the Mosaic
CCDs are unstable at the 2--3 ADU level over a timescale of minutes, so
for all subsequent images, we subtracted both the overscan strip and
the combined bias image for the appropriate night.

Also, for each night we combined the dome flats for each filter into a
single image.  Dividing dome flats from different nights as a test, we
immediately noticed substantial deviations from unity, often in the
form of strong gradients.  Communication with G.~Jacoby led us to
believe that this may have arisen because we were insufficiently
careful when pointing the telescope at the dome white spot when taking
these flats.  The small spot compared to the wide field of the camera
requires pointing to within 1{\arcmin} to prevent large vignetting.
This problem led us to reject the dome flats.

Instead, we used the twilight flats for flat fielding.  First we
selected the twilight flats from each night with mean count levels of
7,000--15,000 ADU (9 in $I$, 10 in $R$ and 15 in $V$).  We combined
these to create one combined flat per filter for the entire run.
Contrary to typical techniques, we did not pre-flatten these with the
dome flats.

We performed initial image reductions on the Landolt fields with
MSCRED-CCDPROC using the twilight flats.  For each field we then found
a single star from the Tycho catalog \citep{esa97} with $V \sim 12$ to
roughly register the {\VRI} images using MSCZERO.  We computed more
detailed astrometry for each image with MSCCMATCH using 500--1000
stars from the USNO-A2.0 catalog \citep{mon96} with $11 < R \lesssim
14.5$.  This typically yielded astrometric solutions with rms errors
of 0.7{\arcsec} to 0.9{\arcsec} in Right Ascension and 0.5{\arcsec} to
0.8{\arcsec} in Declination.  Using these solutions, we resampled the
images to the sky tangent plane with MSCIMAGE.  This latter step was
essential to correct the non-uniform plate scale caused by the field
corrector on the telescope.  Additionally, we found that redoing the
astrometric solution with CCMAP after resampling allows us to improve
the RMS coordinate residuals to about 0.6{\arcsec} in Right Ascension
and 0.45{\arcsec} in Declination.  The greatest improvements occur at
the edges and corners of the image, where the position implied by the
coordinates may move by up to a pixel or two.  We see this improvement
because CCMAP is more flexible than MSCCMATCH, and it appears that
there are minor imperfections in the NOAO-provided plate solution for
the Mosaic field.

Finally, using the reduced Landolt field images we manually identified
and centroided each Landolt star with IMEXAMINE and ran PHOT on those
stars.  We used the relative magnitudes reported by PHOT to test the
photometricity of each night. With one small exception, we found
that Nights 1--3, 5 and 6 are all of high quality with rms fit
residuals of 0.030 ($B$), 0.027 ($V$), 0.027 ($R$) and 0.029 ($I$)
mag.  The one exception is the last hour of Night 1, when thin clouds
appeared unexpectedly.  These clouds ruined the last Landolt field and
damaged the three previous science fields.  The repair of these three
fields is described below in \S~\ref{sec:data-quality}.
As significant difference in the photometric
parameters were not found between the nights, we combined the observations
of standards on all photometric nights to compute the CCD and atmospheric
parameters for the run.  To accomplish the latter, we used FITPARAMS
for each filter to solve for four coefficients: the zero-point, color,
airmass, and color-airmass terms.  

The initial stages of the reduction of the {\VRI} science fields were
identical to those of the Landolt fields.  However, we employed the
additional step of combining the three exposures per filter for the
survey fields.  After resampling the images to the sky tangent plane,
we median-combined the multiple exposures with MSCSTACK to make a
single image.  Because the MSCIMAGE task allows one to register images
while resampling, there was no need for a separate registration step.

Next, we identified the targets for which we wish to compute
photometry.  Rather than attempting to identify sources in our images
with a tool like DAOFIND, we have adopted the point-source detections
from the USNO catalog.  This choice has both positive and negative
consequences.  The main advantage is that we can trust that our
targets are genuine stars and not extended sources, cosmic rays or CCD
artifacts.  In addition, we do not have to attempt the difficult task
of cross-referencing our target detections from the different filters:
we use the same target list for all filters.  The main disadvantage is
that the Palomar survey, on which the USNO catalog is based, has
slightly coarser resolution than the Mosaic, so a close pair which is
resolvable in our field could appear blended as a single star in the
USNO catalog.  Another potential problem is that the USNO coordinates
are epoch 1950--1956, and so may be inaccurate for epoch 1999.  This
concern is minimized because (1) $\lambda$~Ori is very near the
anti-center of the solar motion so solar reflex motion is small, (2)
we suspect relative proper motions of member stars of the association
are quite low, as they are in the Orion Nebula Cluster (most recently
measured as $\mu \simeq 0.08$ mas/yr; \citealt{tia96}), and (3) we are
using large apertures.  Using the astrometric solution from CCMAP
above, we transform the RA/Dec coordinates into pixel coordinates for
all USNO-A2.0 stars with $R < 18$.

We then performed aperture photometry with PHOT on the USNO positions.
Because the $\lambda$~Ori region is not crowded, we used a generous
aperture radius of 12 pixels (about 5{\arcsec}), as we did for the
Landolt stars.  For background subtraction, we used a sky annulus with
inner and outer radius of 18 and 27 pixels respectively.  Following
the PHOT task, we used the {\VRI} portion of the Landolt solution with
the INVERT task to compute calibrated magnitudes for each star.  After
removing the invalid stars (ones flagged by IRAF as saturated or on
the edge of the image), this resulted in photometry of 3000--9000 stars
per field (varying primarily as a function of galactic latitude and
cloud density in the star-forming complex).


In Table~\ref{tab:data} we present $V$, $R$ and $I$ photometry for all
stars, sorted by Right Ascension.  Where we have multiple
measurements, we average the $V$, $R$ and $I$ quantities.  The quoted
errors, derived in the following section, are separated into the three
domains where we are limited by systematics ($12<R<16$), photon noise
($16<R<17$) and sky background ($17<R<18$).  The final three columns
contain J2000.0 coordinates from the USNO-A2.0 catalog, and the
field(s) in which the star was observed.


Figure~\ref{fig:manystars} shows the position of all survey stars with
$11.5 < R < 16.5$.  With these 138,576 stars marked as black dots,
regions of high extinction show up clearly in white, signifying an
absence of background stars.  These regions closely correlate with the
IRAS infrared image of the dense molecular clouds shown in Figure~16
from Paper~II.  Thus Figure~\ref{fig:manystars} vividly demonstrates
that most of the stars in the dataset are background stars.

\subsection{Photometric Quality}
\label{sec:data-quality}

With three exposures per field and overlapping regions between fields,
we have many opportunities for internal quality checks.  Paper~II
discussed in detail the quality of a subset of the photometry,
primarily from Nights 1 and 2.  Here we discuss the entire dataset,
first in a global sense and then followed by an investigation into
differences between fields.


In Paper~II we evaluated our internal errors based on comparisons of
photometry derived from each of the three exposures for each field and
filter.  There, we found an internal limiting precision of 0.016 mag
in all bands for bright stars ($R<15.5$).  For single exposures, stars
with $15.5<R<17$ are photon-limited while stars fainter than $R=17$
are limited by sky brightness.  These limits can be seen in the top
panel of Figure~\ref{fig:photerr} which shows the measured magnitude
differences derived from any two of the three $R$ exposures of field
110.  However, as we will show below, external systematic effects
dominate stars brighter than $R=16$, so the 0.016 mag precision is not
achieved.

Our external errors can be assessed by comparison of photometric
measurements of any star observed more than once.  Such multiple
observation come from fields which overlap each other on edges or
corners, totaling to 44414, 3052, and 399 stars with 2, 3, or 4
observations respectively.  For these stars, we computed standard
deviations of $R$ magnitude measurements.  To study the distribution
of standard deviations, we divided the stars into 6 bins according to
their mean $R$ magnitude.  Inverse Gaussian analysis of each of the
magnitude intervals found the standard deviations of these
distributions to be 0.12 ($18>R>17$), 0.069 ($17>R>16$), 0.047
($16>R>15$), 0.041 ($15>R>14$), 0.039 ($14>R>13$) and 0.037
($13>R>12$) mag.  These numbers demonstrate that photometric quality
is only weakly a function of magnitude for stars with $R<16$.  The
asymptotic approach to a standard deviation of about 0.030 mag
indicates a systematic limit on our photometric precision.  For the
faintest stars, the photometry deteriorates rapidly as the moonlit sky
brightness begins to dominate.  Thus we use the dispersions quoted
above as our photometric uncertainty in Table~\ref{tab:data} for stars
fainter than $R=16$.

However, for stars brighter than $R=16$ this analysis underestimates
the quality of the photometry.  We have found that the field-to-field
overlap regions at the edges of the images (from where all of our
multiple measurements derive) suffer significantly worse photometry
than the rest of the images.  There are two main systematic effects
which degrade the photometric quality.  First, when we compare
photometry between fields which overlap north-south, we find that
there is a systematic offset.  This offset has a mean value of 0.029
($V$), 0.047 ($R$) or 0.051 ($I$) mag measured from all multiply
measured stars with $11<R<16$ in the 64 pairs of north-south adjacent
fields.  These offsets are in the sense that stars at the north edge
of the south-most field appear fainter than the same stars observed in
the south edge of the north-most field.  Measuring this magnitude
difference as a function of position within the overlapping strip, we
find that the most deviant photometry comes from within
$\sim$1{\arcmin} of the very edge of the chip, and improves rapidly
toward the interior of the chip.  We find the dispersion of magnitude
{\em differences} to be 0.048 mag in all three filters, corresponding
to a true measurement precision of a single observation of 0.034 mag.

For the 47 fields which overlap east-west, on the other hand, we find
mean offsets near zero in all filters.  The single measurement
dispersion is 0.030 mag, similar to the difference dispersion found at
the north-south edges.

Given the restriction of the north-south photometric offsets to the
outer couple arcminutes of the field of view and the uniform
photometric quality with position in the east-west overlap regions, we
are confident that photometry over almost all of a field is of uniform
quality.  In the absence of independent photometry, we choose to
minimize the impact of the north-south offset by averaging together
all multiple measurements.  We nonetheless caution that the photometry
of stars within a few arcminutes of the north or south boundary of a
field is likely less accurate than the rest of the stellar sample.

The second effect that systematically degrades the photometric quality is
the proximity of the moon on the later nights of the run.  By Night 6,
the moon was 97\% full and only 18{\arcdeg} northwest of
$\lambda$~Ori.  These conditions forced us to select targets in the
southeast quadrant of the star-forming region until the moon was
occulted by the dome in the last third of the night.  These conditions
inflated the resulting photometric dispersion between overlapping
fields by 15\% or more for stars with $R>16$.

To more accurately quantify the effect of the moon, we recomputed the
photometric precision excluding all fields from Nights 5 and 6, which
translates to excluding almost 50\% of the multiply-measured stars.
The offsets (either north-south or east-west) did not change
appreciably, but the dispersion of the magnitude differences decreased
to 0.030 (north-south) or 0.029 (east-west) mag for a single field.
Comparable tests including {\it just} Nights 5 and 6 yield 0.036 and
0.033 mag.

In conclusion, we adopt an intermediate estimate of 0.032 mag for our
photometric precision in all filters.  The photometry may be worse
near the north and south edges of the fields, but those are regions
where we have multiple measurements whose combination will reduce
the errors.  In addition, we warn that this precision varies from
field to field.  In particular we note that the slightly lower-quality
photometry from later nights of the run are at systematically larger
radii from $\lambda$~Ori.

Finally, we found a few fields to have anomalous photometry.  We
observed six instances (Fields 109, 110, 112, 128, 138 and 173) where
one exposure in one filter had photometry which disagreed with the
other two exposures by between 0.02 and 0.10 mag.  In one additional
case (Field 111), one exposure in each of the three filters suffered
similar defects.  Figure~\ref{fig:photerr} shows data from Field 110
as an example comparing the three exposures in a normal filter
alongside three exposures from a problematic filter.  From this
comparison, one can see the obvious 0.030 magnitude offset in exposure
2 of the $I$ filter.  In all cases, the observed discrepancies were in
the sense that the anomalous exposure generated fainter magnitudes
than the others.  We are not certain about the cause of these
irregularities.  It could be very thin cirrus which we did not notice
at the telescope.  Or, perhaps more likely, it could be due to partial
occultation by the dome, since we had problems with the dome drive
motor overshooting its target azimuth throughout the run.  We have
responded to these problems in the data with two prodedures.  
For Fields 138
and 173, which each had small offsets of about 0.03 mag for one
exposure, we simply applied a correction to that one exposure to match
the others before the coadding step.  A followup comparison with the
overlapping neighbors show that this correction was successful.  For
the rest of the problematic fields, we simply discarded the errant
exposure and combined the other two exposures alone.  Subsequent
checking against the adjacent fields showed that photometry derived
from those remaining exposures is of high quality.  Discarding one
exposure caused us to miss a few stars in the inter-chip gaps in the
affected filter, but the effect is minor.

In addition, we have three fields (Fields 4, 6 and 106, observed in
this order of time) which were apparently contaminated by unnoticed
cirrus at the end of Night 1.  The gradual arrival of these clouds is
obvious when the these fields are compared to adjacent fields observed
on different nights.  The fields surrounding the affected ones all had
very high quality photometry which allowed us to compute precise
offsets.  The first two filters of Field 4 are unaffected while the
third ($I$) suffers 0.015 mag of extinction.  Next, Field 6 suffers
extinction in all three filters of 0.032 ($V$), 0.053 ($R$) and 0.044
($I$) mag.  Finally, Field 106 suffers 0.092 ($V$), 0.144 ($R$) and
0.299 ($I$) mag of extinction.  We used these measured offsets as
corrections which we applied to the final photometry for each of these
fields.  Since the corrections are small (at least for Fields 4 and
6), the errors in the color correction should also be small.  Thus,
our only outstanding concern for these fields is that the extinction
might have been non-uniform.  The photometric offsets for the opposing
edges were similar, so we suspect that the data are of acceptable
quality.

\bigskip

In summary, we have obtained {\VRI} photometric data for 320,917 stars
over 60 degrees$^2$ with a typical photometric precision of 0.032 mag
in all bands for stars brighter than $R=16$.

\section{Search for PMS Stars}
\label{sec:search}

\subsection{Technique}

Our goal is to use these photometric data to identify a statistical
sample of PMS stars in the $\lambda$~Ori star-forming region.  We have
demonstrated in Papers~I and II that using spectroscopic data combined
with multicolor photometry we can securely identify young association
members despite the fact that field stars outnumber the members by
factors of several hundred.  In this section, we demonstrate that we
can use the knowledge gained from our spectroscopic survey to identify
a statistical sample of stars which closely resemble the PMS
population (in number, spatial distribution, and photometry) using the
photometric data alone.  As a proxy for the true PMS population, we
will use this sample to study the spatial distribution of young,
low-mass stars over the entire $\lambda$~Ori region and to compute the
IMF of the star-forming complex.

Our technique relies on the fact that PMS stars have $R$ vs.\ {\RI}
CMD positions rather different from most field stars.  Additionally,
we use the information from Papers~I and II that the PMS stars lie in
a small range of ages at about the same distance.  Exploiting these
characteristics allows us to maximize PMS stars while minimizing field
stars in a photometrically-selected sample.  We use the $R$, {\RI} and
{\VR} measurements of confirmed PMS stars to define photometric
boundaries in which the field star contamination will be as low as
possible.

\subsection{Calibration}



The first step is to define a region of the CMD which has a high ratio
of PMS stars to field stars.  We use our spectroscopic PMS population
from Paper~II as a calibration sample, or control group, to define
such a region before applying it to the fields where we have no
spectroscopic information.  Figure~\ref{fig:control} shows the
location of these spectroscopically-identified PMS stars along with a
large population of field stars in the same lines of sight.  Using a line at
$R=16$ along with isochrones and mass tracks from the \citet{pal99}
stellar evolution model as boundaries, we can isolate a region in the
CMD which has many PMS stars and few field stars.  Specifically, we
have selected an isochrone at 4 Myr and a mass track at 0.6 {\msun}
(both at 450 pc) as optimal boundaries to minimize field
contamination.


We further cull field stars from the spectroscopic control sample
using the {\vr} colors.  Figure~\ref{fig:cullvr} shows all of the
field and PMS stars which fulfill the criteria in the $R$ vs.\ {\RI}
domain.  For this sub-sample, the PMS stars form a well-defined locus
in the $R$ vs.\ {\VR} CMD, while the field stars are much more broadly
distributed, particularly brighter and redder than the PMS stars.  We
have chosen a pair of boundaries, shown in the figure, which remove 28\% of
the field stars at the expense of only 1\% of the PMS stars.  The
resulting {\VR}, {\RI} and $R$ boundaries 
yields roughly equal numbers of PMS (179) and field (232) stars.

\subsection{Application to Survey}

Next we use these photometric boundaries to select candidate PMS stars
in survey fields for which we do not have spectroscopic information.
Figure~\ref{fig:agemap} shows the spatial distribution of all stars
which fall within the photometric criteria defined above.  Inset in
that figure is an $R$ vs.\ {\RI} CMD showing those same stars.
Comparison with Figure~\ref{fig:manystars} gives a sense of the large
extent to which field stars have been removed.


The most obvious feature of this map is that there are stars
everywhere in the region of interest.  Based on the analysis of the
previous section, more than half of these are field stars which should
have a nearly uniform distribution.  Close examination of
Figure~\ref{fig:agemap} shows that there are excesses of stars in
particular areas.  Notably, there is a large excess centered on
$\lambda$~Ori.  There is also a distinct clustering of stars near B35.
Thus, the photometric selection technique is successfully identifying
PMS populations in regions where we had previously found enhancements
of PMS stars by spectroscopic techniques.

Additionally, the figure appears to show a ring of higher stellar
density at the edges of the surveyed region, coinciding with areas
where the field star density is shown to be low in
Figure~\ref{fig:manystars}.  However, as we discuss in the next
section, these density enhancements may be artifacts caused by
reddening in the dark clouds rather than enhancements of PMS stars.

\subsection{The Effects of Extinction}
\label{sec:extinction}

When searching for density enhancements in Figure~\ref{fig:agemap},
one implicitly assumes that the field star distribution is uniform.
However, if a large fraction of the field stars are more distant than
the $\lambda$~Ori association as indicated by
Figure~\ref{fig:manystars}, the reddening due to dense gas in the
star-forming region can move field stars into our photometric
selection region.  This change in the morphology of the field CMD can
cause an apparent excess in the vicinity of a dusty cloud.  That is,
the number of non-PMS stars varies as a function of extinction in the
star-forming region.

To estimate the magnitude of the extinction, we use the following
simple procedure.  First we refer back to Figure~2 of Paper~II, which
shows the distribution of CO J=1$\rightarrow$0 emission in the
star-forming region \citep{lan98}.  The highest contour is 19.2 K
{\kms}.  \citet{lan98} quote a conversion from CO brightness
temperature to H$_2$ column density of $(1.06\pm0.14)\times10^{20}{\rm
cm}^{-2}$ (K {\kms}) from \citet{dig95} for Orion.  Thus the strongest
CO emission traces H$_2$ of column density $2\times10^{21}$ cm$^{-2}$.
Doubling this number to account for the two hydrogen atoms per
molecule and using the \citet{boh78} conversion from hydrogen column
density to $E_{B-V}$ color excess of $5.8\times 10^{21}$ cm$^{-2}$
mag$^{-1}$, we find the peak reddening is roughly $E_{B-V}=0.7$ mag.
Using a general-to-selective extinction ratio of $R_V=3.1$ and the
\citet{car89} extinction law, we find the peak absorption to be on the
order of $A_R=1.6$ mag.  If instead we use a higher value of $R_V=5$,
we find $A_R=2.7$ mag.  Certainly, this calculation is only an
approximation (for example it is not clear if our application of the
Bohlin, Savage \& Drake conversion is appropriate for this region.
However
it suffices to demonstrate that the dense clouds have non-negligible
extinction, but are not opaque.

Thus, there could be a couple of magnitudes of extinction for
background stars where the molecular clouds are densest.  The
consequence of this is that many of the blue field stars could be
reddened into the photometric boundaries we defined above.

To test this hypothesis, we take a field where there is very little
molecular gas (as inferred from the CO map) and artificially redden it
by a constant extinction, adding some random scatter to simulate varying
conditions within the field.  Figure~\ref{fig:redtest} shows an
example of this test performed on Field 151.  The upper left panel
shows the unaltered CMD of that field.  To create the CMD in the lower
left panel, we added an extinction ($A_R$)  randomly selected from
a gaussian distribution with a mean of 1 magnitude and a standard deviation
of 0.5 mag (discarding any selected extinction below zero).  
In this example we compute the $E_{R-I}$ reddening
from $A_R$ using $R_V = 3.1$.  As a comparison, we include in the
lower right panel the CMD of Field 65, which should be genuinely
reddened as it is projected on the B30 dark cloud where the CO has a
high column density.


Note the strong similarity between the lower two panels of
Figure~\ref{fig:redtest} showing one artificially and one physically
extincted field.  We take this similarity to mean that (1) the on-cloud
fields have similar background 
CMDs to the off-cloud fields, but they have been extincted and
reddened; (2) most of the field stars are in the background; and (3)
our extinction estimates calculated above are reasonable.

Most importantly, the number of stars in the photometric selection
region rises rapidly with increasing extinction.  Therefore, we
conclude that the apparent excess stellar density projected on the
regions of the dark clouds in Figure~\ref{fig:agemap} is likely an
artifact of the extinction of background field stars caused by those
clouds.  This also means that the probability of a bright, red star
projected on a dark cloud being a young member of the association is
much smaller than one with similar photometry projected on a
transparent region.

\subsection{Statistical Removal of Field Stars}
\label{sec:search-remove}

The sample of stars we constructed above contains stars with
photometry consistent with PMS members of the association.  However,
that sample still contains a substantial number of field stars which
are photometrically indistinguishable from PMS stars in {\VRI}.  To
study the IMF and spatial distribution of the low-mass $\lambda$~Ori
members, we would like our proxy population to contain stars with not
only the correct photometric properties (thus ages and masses), but
also to contain the correct number of stars.  Ideally, we would simply
subtract a constant number of stars from each field to arrive at a
statistical estimate of the number of PMS stars in those fields, but
unfortunately the field star distribution is not uniform, 
because the background star density decreases with increasing galactic
latitude.  Therefore, in this section we describe a technique we have
used to determine the number of field stars using the assumption that
the unreddened field-star CMD has the same morphology across the
entire complex.  That is, we assume that the ratio of field stars in
one region of the CMD to another region is a constant.  We
establish this ratio using the spectroscopically studied 
control fields where we can securely create a
sample with only field stars and no PMS stars.  Then we use this
ratio to subtract field stars in each of the rest of the fields.  This
yields a measure of the number of PMS stars in each field.

Note that, as described in the last section, the regions with
substantial reddening do not have the same CMD morphology as
unreddened regions, so this technique does not work well for fields
projected on the dark clouds.

In detail, we first identified a region of the CMD which is free of
PMS stars (which we call the ``PMS-free region'' in contrast to the
``PMS candidate region'' discussed in earlier sections).  We defined
this by the boundaries $0.5 \leq R-I \leq 0.7$ and $14.5 \leq R \leq
15.5$.  Several criteria entered into our selection of this particular
region.  First, this region is rich with field stars, as seen in
Figure~\ref{fig:control}, for example.  The more field stars there
are, the less error is introduced into our calculations by counting
noise.  Second, this region is faint enough and blue enough that it
cannot contain any PMS stars.  In fact, this box is almost entirely
below the ZAMS at 450 pc, let alone near the PMS stars.  Third, it is
bright enough that photometric is negligible.

Totaling over all the control fields except those with poor
spectroscopic completeness (as identified in Table~3 of Paper~II) and
those projected on the densest portions of B30, we find a ratio of
0.033 for the number of field stars\footnote{That is, the number of
spectroscopically identified field stars corrected for incompleteness.}
in the PMS-candidate zone to the number of stars in the PMS-free
region.  The counting error associated with this ratio is about 15\%.

Next, for each of the fields in our survey we count the number of
stars in the PMS-free region and use the above ratio to compute the
expected number of field stars in the PMS-candidate region.  The
difference between this result and the observed number of PMS
candidates is an unbiased estimate of the true number of PMS stars.  

Finally, for the purpose of creating a map of the PMS stars, we
subtract the derived number of 
field stars from the survey fields.  This removal is
statistical, since we do not really know which are field stars and
which are PMS stars.  Thus, for each field we randomly remove a number
of stars equal to the expected number of field stars leaving a
representative sample of PMS candidates with approximately the same
numbers, photometric properties, and spatial distribution of the true
PMS population.

The random nature of the removal technique means that the PMS
sample is only representative on a field-by-field basis.  That is, we find
an accurate estimate of the number of stars in each 1{\arcdeg} field,
but we do not know their true positions within that field.  Because
the fields which are projected on the molecular clouds suffer
reddening of the background stars, our technique is invalid
there\footnote{Note that just because our technique does not work well
on the cloud regions does not mean that there are no PMS stars there.  On
the contrary, our spectroscopic survey detected substantial numbers of
PMS stars projected on the B30 cloud (in the upper right portion of
the figure), but these are outnumbered by the reddened field stars
masquerading as PMS stars in the figure.} (although it has been 
applied to
all fields for simplicity).

Figure~\ref{fig:pmsmap} shows the statistical PMS population after
subtraction of field stars.  Figure~\ref{fig:pmsmap2} shows the same
map as Figure~\ref{fig:pmsmap} rotated into galactic coordinates and
overlayed on the CO map of \citet{lan98}.  Stellar density
enhancements are clearly seen near both $\lambda$~Ori and B35,
confirming similar observations made from Figure~\ref{fig:agemap}.
{\em Perhaps more remarkably, much of the region inside the molecular
ring is devoid of PMS stars.}



These two figures show a large number of low-mass stars near
$\lambda$~Ori and the neighboring B stars.  The stars around
$\lambda$~Ori are present primarily within the 2{\arcdeg} circle
marked on Figure~\ref{fig:pmsmap}.  Out to that radius, this
distribution is consistent with an $r^{-2}$ density distribution
centered on $\lambda$~Ori.  This is reminiscent of the simple
kinematic model of the stellar expansion we constructed in Paper~I
where we explored (and could not rule out) the possibility that all of
the stars around $\lambda$~Ori had formed in a tight Trapezium-like
cluster and then kinematically dispersed when the parent gas was
dispersed.

The essential new information from this study is the spatial extent of
the population of PMS stars around $\lambda$~Ori.  Even excluding
stars in the vicinity of B35, the distribution of PMS stars extends as
far as 16 pc from $\lambda$~Ori.  For a one-dimensional velocity
dispersion of 2.5 km/sec (as found from radial velocities in Paper~I),
a typical star would move 2.5--5 pc in 1--2 Myr, while extreme
velocity stars might travel as much as 7.5--15 pc.  As such it is
possible that the spatial extent of the present PMS population around
$\lambda$~Ori derives in large part from ballistic expansion of an
initially more concentrated population after a recent gas dispersal
event.

Even so, given the substantial fraction of stars found 10--15
pc from $\lambda$~Ori, it seems unlikely that the distributed
population derives entirely from a Trapezium-like cluster around
$\lambda$~Ori.  Indeed, as noted in Paper~I, it would be odd if the
low-mass stars of such a cluster dispersed while the OB stars did not.
Thus we conclude that star formation occurred over a distributed
region around $\lambda$~Ori, but quite possibly over a spatial extent
more limited than the presently observed spatial distribution of PMS
stars.

The edge of this central population of PMS stars abuts on a clump of
stars near the B35 cloud.  In Figure~\ref{fig:pmsmap2}, one can see
that this group extends toward $\lambda$~Ori from the densest portion
of the molecular cloud.  However, it is somewhat surprising that there
appear to be no PMS candidates on the ``neck'' of the B35 cloud, the
elongated portion which connects the dense ``head'' to the ring.
Clearly the star-formation rate has differed significantly between
the front and the back of this cloud.

Finally, the {\it lack} of PMS candidates outside of the central
population is equally important.  Figure~\ref{fig:pmsmap} shows this
dearth most clearly in the region between the two circles.  (Again,
the surface density enhancements outside the larger circle are not valid
representations of the PMS population.)  Thus, within the present ring
of molecular clouds we find that star formation occurred only within a
2{\arcdeg}-radius circle around $\lambda$~Ori and near the B35 cloud.
We also know from Paper~II that star formation has been active in the
B30 cloud.  It remains to be seen whether any star formation has
occurred elsewhere within the molecular ring.

\subsection{The Initial Mass Function}

Using our statistical count of PMS stars to represent the true PMS
population, we assess the association IMF in this section.  We
restrict our sample to $0.4 M_\odot < M < 0.6 M_\odot$, which
minimizes both incompleteness at the faint end (see \S~5.3 of
Paper~II) and contamination by the field, respectively.  In the
census completed in Section 3.5, we used an age limit of 4 Myr.  However, in Paper~II
we noted that the onset of star formation occurred about 8 Myr ago.
Thus a fair assessment of the IMF should include stars of at
least that age.  A drawback is that in extending our census 
to an 8 Myr age limit, the number of field stars
increases faster than the number of PMS stars. For stars with
with $0.4 M_\odot < M < 0.6 M_\odot$ there
are 60\% more
total stars in the expanded selection region of the CMD, but in the
spectroscopic survey we found only a 10\% increase in PMS stars in this
region (62 PMS stars younger
than 8 Myr vs.\ 56 younger than 4 Myr).

Thus we consider two different methods of counting PMS stars to 8 Myr.
First, we simply redo the census with an age cutoff of 8 Myr (which involves
re-doing the field star calibration and rejection of
\S~\ref{sec:search-remove}).  In this case, we count 142 stars 
with $0.4 M_\odot < M < 0.6 M_\odot$
within the dashed, 3{\arcdeg} circle on
Figure~\ref{fig:pmsmap}.
Second, we consider the number of PMS
candidates younger than 4 Myr (102) and extrapolate to 8 Myr using the
ratio of PMS stars younger than 8
Myr (62) to those younger than 4 Myr (56) found in our spectroscopic survey. 
This yields an estimate of
(102$\times$62/56) = 113 PMS stars younger than 8 Myr over the
entire star-forming region.  This latter computation has the drawback
that it implicitly assumes the same star-formation history throughout
the entire region.  Since we do not know which count is more accurate,
we will use both the direct count (142 PMS stars) and the extrapolated
count (113 PMS stars) in the calculations below.
We note that our uncertainty in the calibration of the field star
density of 15\% yields a 20\% uncertainty in the number of PMS
candidates.  

Using the \citet[hereafter MS]{mil79} field IMF, 142 stars 
with $0.4 M_\odot < M < 0.6 M_\odot$ predicts 44 OB
stars; a count of 113 PMS stars predicts 35 OB stars.
In fact, there are only 24 OB stars (as identified in Paper~II via the
PPM catalog; \citealt{roe88}) in projection within the region of the low-mass
star census.  Application of the
Monte Carlo test described in Paper~I shows that the difference
between the observed number and the prediction suggests a difference in
the IMFs at the 99.8\% confidence level
for the direct count and at the 95.8\% confidence level for the
corrected count.  Thus, according to the MS IMF, the low-mass stars
are mildly over-represented in the star-forming region.  This
over-representation is further increased by the fact that there are very likely
non-members included in the OB census, as discussed in \S~3 of
Paper~II.

We have also tested the field IMF formulation of \citet[hereafter
KTG]{kro93}, which employs three power-law fits to the mass
function instead of the log-Gaussian used by MS.  The KTG field IMF
predicts that, given the direct count of 142 stars between 0.4 and 0.6
{\msun}, we should see 20 OB stars.  This is more than a factor of two
lower than the MS prediction above, and is consistent with the
observed number of OB stars.  The corrected count of 113 PMS stars
predicts 16 OB stars which is also consistent with the observed number.

Clearly the systematic uncertainties in the field IMF dominate these
comparisons. Given that only in the most extreme case can the difference between
the field IMF and the association IMF be considered significant, we 
conclude that the global initial mass
function in $\lambda$~Ori is not distinguishable from that in the field.

On the other hand, local variations of the IMF are significant, as reported in
Papers~I and II: (1) in the central 1{\arcdeg} where the OB stars are
concentrated, the low-mass stars are {\it deficient} by a factor of
two; (2) outside of this central region they are {\it overabundant} by
a factor of three.  The spatial distribution of star formation is
significantly {\it mass-biased}.

Our count of stars with $0.4 M_\odot < M < 0.6
M_\odot$ plus the 24 OB stars implies a total mass of 
450 to 650 {\msun} for stars above 0.1 {\msun} within the
molecular ring, depending on the PMS count (113 or 142) and choice of
field model (MS or KTG).  These mass estimates are a factor of two
higher than the the lower limit on the total mass stated at the end of
Paper~II, but here we also include more than a factor of
three more spatial coverage.  Compared to the total cloud mass of
$4\times 10^4$ {\msun} calculated by \citet{mad87} (including
molecular, neutral and ionized gas), this implies a star formation
efficiency of 1\%--2\%.  The molecular cloud mass computed with higher
quality CO data by \citet{lan98} matches that of Maddalena \&
Morris for the molecular gas ($1\times 10^4$ {\msun}), but they do not
include \ion{H}{1} or \ion{H}{2} mass, so we cannot compute an
independent efficiency from their work.  One must note that these are
mass estimates of the present-day clouds.  It is likely that a
significant fraction of the natal cloud could be sufficiently heated
or dispersed by now that it is not included in the above estimates.
But whether the efficiency is 1\%--2\% or a factor of two lower, it is
similar to other estimates of global 
star-formation efficiency in molecular clouds 
(e.g.\
\citealt{mye86}).

\section{Interpretation: The Supernova Scenario Revisited}

We have created a statistical representation of the entire young
stellar population inside of the molecular ring.  We have found a
large concentration of PMS candidates centered on $\lambda$~Ori as
well a concentration near the B35 dark cloud.  More thorough studies
of a subset of these stars via WIYN spectroscopy in Papers~I and II
definitively showed that these populations are composed of genuine
young members.  In addition, the spectroscopic investigation showed
that there are many PMS stars projected on the B30 dark cloud (on the
northwest side of the molecular ring), where our purely photometric
analysis does not work well.  There are no other concentrations of
young stars within the molecular ring.

This supports the model proposed by \citet{due82} that the initial
cloud distribution was elongated, extending (at least) from B30
through the center to B35.  Thus, the current distribution of young
stars is seen as a fossil of the parent molecular cloud, tracing the
densest parts of that structure.  Our identification of this fossil is
only possible because (1) the cloud is mostly dispersed today and (2)
the stars have not moved much from their places of birth.  This latter
point is supported by the kinematic evidence presented in \S~3.4 and
\S~4.4 of Paper~I.  To recap those arguments, the high OB-star proper
motions and PMS-star radial velocities in combination with the present
tight spatial concentration of those populations constrain the
dynamical timescale to 1--2 Myr, at most.  Before that, the stars must
have been gravitationally bound, presumably by the parent cloud.

But today there are virtually no signs of that parent cloud remaining
within the central 2{\arcdeg} radius, or 16 pc radius at the 450 pc
distance of $\lambda$~Ori.  We have suggested that the cloud was
rapidly dispersed by a supernova about 1 Myr ago.  We require a
supernova instead of more conventional action by OB star winds and
radiation primarily to achieve the short timescale: \citet{mad87}
calculate that more than 3 Myr is needed for the OB stars alone to
carve out an {\hii} region of 16 pc radius.  In 3 Myr, the proper
motion of those OB stars would have carried them outside of the
molecular ring, which is clearly inconsistent with observations.

In addition, the dearth of PMS stars younger than 1--2 Myr within 16
pc of $\lambda$~Ori (discussed in detail in Paper~II), despite the
abundance of such young stars in the more distant dark clouds B30 and
B35, demonstrates that an event centered near $\lambda$~Ori terminated
all low-mass star formation in its vicinity quite recently.  The
probable mechanism for this termination is the dispersal (again by
supernova) of the central cloud from which the stars were forming.



With this evidence as foundation, we construct the following 
history of the star-forming region.  In the course
of this discussion, the reader may wish to refer to Panels A, B, C and
D of Figures~\ref{fig:schemzoom} and \ref{fig:schematic}, which
show schematic representations of this history.

About 8--10 Myr ago, the $\lambda$~Ori region was composed of a
starless, roughly linear string of dense molecular clouds (Panel A).
The most massive lobes of this cloud chain were the large central
core, the progenitors of the present-day B30 and B35 dark clouds, and
(south of these) the B223 cloud.  The elongated structure of this
cloud complex connected to the other linear structures to the
northwest and southeast which we see today in the CO map shown in
Panel D of Figure~\ref{fig:schematic}.

Over the next few Myr, stars began to form in the densest portions of
this cloud chain.  At 6 Myr ago, a dozen OB stars formed near
$\lambda$~Ori's present-day position (Panel B).  The local gas density
remained high enough that they could not ionize or disperse more than
a small fraction of the gas.  Thus, they remained bound to their natal
cloud.  At the same time, the birth rate of low-mass stars increased
in all productive areas of the star-forming complex.  Many of these
were unaffected by the massive stars many parsecs away, but the
closest PMS stars passed in close proximity to the OB stars and lost
their circumstellar disks to the FUV radiation (Paper~I).

Then, shortly before the present day, one of the O stars (perhaps a
binary companion of $\lambda$~Ori itself) became a supernova (Panel
C).  The supernova blast encountered a non-uniform medium filled with
a small {\hii} region forming around the OB stars, dense gas in the
immediate vicinity of the OB stars, a few massive clouds about 10--15
parsecs away, and the rest of the volume filled by somewhat lower
density molecular gas.  The blast quickly dispersed all of the parent
core, creating the molecular ring, the large {\hii} region, and the 
nearby \ion{H}{1} structures
\citep{zha91}.  However, when the shock reached the more distant, massive B30
and B35 clouds it swept around them.
Thus today we see the fossil distribution of
young stars within the molecular ring, as well as
the remnants of the B30 and B35 clouds within the
ionized region (Panel D).

In this scenario we suggest that much of the gas from the parent core has moved
as a consequence of the supernova, while the other massive clouds have moved
little.
This picture differs from the all-expanding
model of \citet{mad87} and the all-stationary model of Paper~II.  The
Maddalena \& Morris model is contradicted because the accelerating gas
should have left behind all stars formed therein, whereas we find
stars of all ages still projected on B30 and B35 (implying that these
clouds have not moved since the oldest stars were born).  Also
incorrect is our Paper~II model of the star-forming region as a single
giant cloud with a hole torn from its center, because we have found an
annulus inside the CO ring absent of PMS stars.  This absence implies
that the annulus never had clouds of high enough density to form
stars.

Examining Figure~\ref{fig:pmsmap}, one can see the CO ring is not actually
centered on $\lambda$~Ori.  We interpret this as evidence of
non-uniform gas density: the supernova shock expanded more easily
towards (and around) the smaller B35 cloud to the east than towards
B30 to the northwest.  Over several hundred thousand years, the
supernova shock snowplowed the gas outward to where we see the cold
ring today.  Because the stars have a radial velocity similar to the
B30/35 clouds (comparing our data to the \citet{lan98} and
\citet{mad86} CO data) while the {\hii} gas has a mean velocity
several {\kms} blueward (from the Wisconsin H$\alpha$ Mapper survey;
L.~M.\ Haffner 1998, private communication), it seems the shock must
have expanded more readily into lower density gas in the foreground.

Certainly, the B30 and B35 clouds have not been totally immune to the
shock wave.  The inward-facing sides of the clouds have been heavily
eroded by the blast.  We see a signature of this in the strong density
gradients in the molecular clouds facing towards $\lambda$~Ori, seen
in Figure~\ref{fig:pmsmap2}.  This is a consequence of either
compression by the supernova (and/or subsequent OB winds) or by the
dispersal of the outer, low-density envelope of the cloud.  These same
faces show bright rims today \citep{lad76} as they are continuing to
be ionized by the remaining OB stars.  We also see evidence of the
partial destruction of B30 and B35 in the PMS stars which lie slightly
in front of the clouds, tracing the former extent of the gas.  We
reiterate that because the stars in front of the clouds have a wide
range of ages matching the age span of the entire association, this
enhanced PMS population in front of the clouds is likely a sign of
cloud destruction, not cloud
acceleration or triggered star formation.

It appears that the supernova had a more devastating effect on B35
than B30, since the former appears to have ceased embedded star
formation while the latter still contains many far-IR sources
\citep{mat90}.  All of the PMS candidates near B35 in
Figure~\ref{fig:pmsmap2} lie on the side of the cloud facing
$\lambda$~Ori; none lie on the low-surface-density ``neck'' connecting
B35 to the CO ring.  We suspect that the B35 cloud will vanish in a
few Myr, especially if $\lambda$~Ori becomes a supernova soon.

There are still enigmas left in the structure of the gas which are not
explained fully by our scenario.  First is the history of the B223
dark cloud in the southwest of the star-forming complex.
\citet{mad87} noted that B223 is blueshifted relative to the rest of
the clouds and interpreted this as evidence of expansion of the gas.
However, we have found that all of the PMS stars are at a similar
radial velocity to B30 and B35, not at a system velocity intermediate
between B30, B35 and B223 as would be expected if all clouds were
expanding from a common center.  The CO surface density for the B223
cloud is quite high, such that its mass is likely comparable to that
of B30.  Thus, it seems likely that if B30 was unmoved by the
supernova, B223 should be similarly stationary.  The radial-velocity
difference of B223 suggests that
either (1) B223 was not originally connected to the $\lambda$~Ori
progenitor cloud, or (2) B223 is the result of the projection or
collision of two clouds: one ejected from the $\lambda$~Ori vicinity
and a denser, unrelated one.  This latter scenario is appealing, since
the velocity profile presented by \citet{lan98} shows that B223 spans
almost 10 {\kms} in radial velocity, which could be a sign of two
superimposed clouds.

Another puzzle is the presence of a pair of low-surface-density CO
clouds projected very near to $\lambda$~Ori
(Figure~\ref{fig:pmsmap2}).  If these clouds are at the same distance
as the OB stars, then it is unclear how they could have survived the
massive star radiation, let alone a supernova.  Instead, we suspect
that these thin clouds must be in front of or behind the stellar
association.  A detailed velocity map with the sensitivity of the
\citet{lan98} survey might help to solve this dilemma.

The scenario we have described includes stars forming near the
massive, stationary clouds, but not elsewhere.  In the vicinity of the
CO ring, where we cannot photometrically sift the PMS candidates from
the field stars, we do not know how many members may exist.  Our model
of expanding, low-mass cloudlets implies that there should be no old
stars in the majority of the ring.  A detailed spectroscopic survey in
the vicinity of these clouds would answer the question of whether star
formation is occurring there or not.  We suspect that there could be
more stars forming just outside the ring, south of Betelgeuse, where
the $\lambda$~Ori region reaches the ``Northern Filament'' identified
by \citet{mad86}.

The total mass of stars identified as candidate members is roughly
500 {\msun}, which is comparable to an open cluster, but spread
over a much larger area (about 30--40 pc diameter).  The kinematics
indicate that this association will be completely dispersed into the
field within a few $\times$ 10 Myr.  Thus, the $\lambda$~Ori region
may be representative of the process of star-formation in
moderate-mass molecular clouds.

\section{Summary}

We have presented photometry of over 300,000 stars in the
$\lambda$~Ori region.  Using the spectroscopic survey from Papers~I
and II to define the photometric characteristics of PMS stars, we
culled more than 99.9\% of the field stars from our sample.  Further
statistical field star subtraction then leaves a representative sample
of PMS candidates which has the same spatial distribution and
population size as the true PMS stars in the association.  This sample
presented us with a representative snapshot of the present-day
low-mass young stellar population of the star-forming complex.  In
combination with the more detailed, but spatially-limited, results
from Papers~I and II, this snapshot allowed us to deduce the
chronology of the region:
\begin{itemize}
\item 10 Myr ago --- A long chain of molecular gas extended from east
to west across the present-day star-forming region, including three
particularly massive clouds.

\item 6 Myr ago --- Stars formed in the most massive clouds, but the
onset of formation was not sudden.  Instead, the birth rate increased
gradually over many million years.  Numerous OB stars were born in the
central cloud, but were very rare elsewhere.

\item 1 Myr ago --- A supernova exploded, shredding the central cloud
and thus unbinding the central stellar population.  A ring of gas was
pushed from the center region.

\item Today --- Star formation continues in the B30 and B35 clouds,
but has ceased in the vicinity of the supernova epicenter.

\item Future --- The termination of star birth in B35 is imminent, as
is the escape of the OB stars from their central position.  All of the
stars will disperse into the field over the next 10 Myr or
so. The gas may also be further dispersed by subsequent supernovae.

\end{itemize}
We have found that globally this star-forming region has generated a
mass distribution similar to the field population, but the IMF is spatially
non-homogeneous: the center region strongly favors massive stars,
while the periphery is heavily biased toward low-mass stars.

If the $\lambda$~Ori region is typical of star formation in
medium-mass molecular clouds, then this history tells us that these
star/cloud associations are short-lived: they terminate themselves
from within and are not detectable for more than a few tens of Myr.
It is not clear what signals the commencement or acceleration of star
birth: in the $\lambda$~Ori region, stars began to form at a low rate
everywhere at about the same time, but most of the stars were formed
recently.  It would be very enlightening to identify analogues of the
$\lambda$~Ori region at earlier and later stages of evolution to study
the turning points of star formation history.

\acknowledgments

This work was supported by NSF grant AST 94-1715 and NASA ADP grant
NRA-98-03-ADP-003.  We are grateful to J.\ Mathis for assistance with
\S~\ref{sec:extinction}.

\figcaption[Dolan.fig1.eps]{Map of the {\VRI} KPNO Mosaic fields
observed in December 1999 (with internal identification numbers).  The
fields are shaded according to the night on which they were observed.
\label{fig:nightmap}}

\figcaption[Dolan.fig2.eps]{Map of all stars with $11.5 < R < 16.5$.  The
white areas show regions of high extinction associated with the ring
of molecular clouds around $\lambda$~Ori.
\label{fig:manystars}}

\figcaption[Dolan.fig3.eps]{Demonstration of normal versus defective
photometry.  The plots show the difference of measured photometry from
two exposures of the same field as a function of magnitude.  Each row
shows the three permutations of the difference between two of the
three exposures taken per filter.  The top row shows an excellent set
of exposures while the bottom demonstrates a case where exposure
number 2 generates photometry 0.030 mag fainter than the other two
exposures.
\label{fig:photerr}}

\figcaption[Dolan.fig4.eps]{$R$ vs.\ {\RI} photometric discrimination of
PMS stars from the field.  The small dots are all the stars in the
fields studied spectroscopically.  The large dots are the PMS stars
found in those same fields.  We overlay a line at $R=16$ along with a
4 Myr isochrone and a 0.6 {\msun} evolution track from the Palla \&
Stahler (1999) model to indicate boundaries used to isolate the
candidate PMS in the upper right.
\label{fig:control}}

\figcaption[Dolan.fig5.eps]{$R$ vs.\ {\VR} photometric discrimination of
PMS stars from the field.  The stars shown are those isolated in the
upper right of Figure~\ref{fig:control}.  The lines are the boundaries
we derive from this figure for use in isolating PMS stars in 
the survey fields.  The lower
line passes through the points ({\RI}, $R$) = (0.4,12) and (1,16)
while the upper line is defined by (0.85,12) and (1.55,16).
\label{fig:cullvr}}

\figcaption[Dolan.fig6.eps]{Map of all stars selected by the photometric
criteria defined in the text (see Figures~\ref{fig:control} and
\ref{fig:cullvr}).  The inset plot shows $R$ vs.\ {\RI} for those same
stars.
\label{fig:agemap}}

\figcaption[Dolan.fig7.eps]{Demonstration of the effects of reddening due
to the dark clouds.  We transform the upper left panel to the lower
left panel by adding a normally-distributed random amount of reddening
with mean and standard deviation of 1 mag and 0.5 mag, respectively.  The lower
right panel shows the observed CMD of 
a field projected on one of the dark clouds as a
comparison.
\label{fig:redtest}}

\figcaption[Dolan.fig8.eps]{Map of PMS candidates with field stars removed
statistically.  The fields outside the dotted circle are projected on
the molecular clouds, so the high surface density of PMS candidates is
very likely artificially enhanced by reddening.  The dotted circle is
centered about 40{\arcmin} southeast of $\lambda$~Ori, and has a
radius of 3{\arcdeg}.  The solid circle is the 2{\arcdeg} radius
boundary centered on $\lambda$~Ori which we use in Paper~II,
and which is shown in Figure~2 of Paper~II.
\label{fig:pmsmap}}

\figcaption[Dolan.fig9.eps]{Same as Figure~\ref{fig:pmsmap}, rotated
and overlayed on the map of \citet{lan98} in galactic coordinates.
\label{fig:pmsmap2}}

\figcaption[Dolan.fig10.eps]{Schematic history of the star-forming
region showing conditions at 10, 6 and 1 Myr ago as well as a map of
the clouds today with CO contours from \citet{mad86}.  
\label{fig:schemzoom}}

\figcaption[Dolan.fig11.eps]{Expanded scale version of
Figure~\ref{fig:schemzoom}, showing the environment around the
$\lambda$~Ori region.  Again, the CO map is from \citet{mad86}.
\label{fig:schematic}}

\begin{deluxetable}{ccccl}
\tablewidth{0pt}
\tablecaption{Observing Summary \label{tab:obs}}
\tablehead{ 
  \colhead{Night} &
  \colhead{Date} &
  \colhead{Fields} &
  \colhead{Comments}
  \\
  &
  &
  \colhead{\VRI} &
  &
}
\startdata
1 & Dec 15 &    11 & Very good, then cirrus just before dawn \\
2 & Dec 16 &    14 & Cirrus at dusk, then very good \\
3 & Dec 17 &    13 & Perfect \\
4 & Dec 18 & \phn0 & Thick clouds, no data \\
5 & Dec 19 & \phn6 & Cirrus early, then clear with bright moon \\
6 & Dec 20 &    17 & Full moon, no clouds \\
\enddata
\end{deluxetable}

\begin{deluxetable}{cccc}
\tablewidth{0pt}
\tablecaption{Field Name Cross Reference \label{tab:chxref}}
\tablehead{ 
  \colhead{Paper~II} &
  \colhead{This paper} &
  \colhead{Paper~II} &
  \colhead{This paper}
}
\startdata
 1 & 107     & 7 & 33 \cr
 2 & 4 and 6 & 8 & 35 \cr
 3 & 4       & 9 & 51 \cr
 4 & 14      & 10 & 53 \cr
 5 & 12      & 11 & 65 \cr
 6 & 11      & & \cr
\enddata
\end{deluxetable}

\begin{deluxetable}{cccc}
\tablewidth{0pt}
\tablecaption{Target Fields \label{tab:fields}}
\tablehead{ 
  \colhead{Field} &
  \colhead{R.A.} &
  \colhead{Decl.} &
  \colhead{Night}
}
\startdata
\phn4 & 05 43 23.01 & 08 56 20.93 & 1 \\
\phn6 & 05 46 53.02 & 08 56 03.43 & 1 \\
11 & 05 35 33.56 & 09 55 13.14 & 1 \\
12 & 05 39 02.40 & 09 55 50.89 & 1 \\
14 & 05 43 00.89 & 09 54 46.50 & 1 \\
\enddata
\tablecomments{The final number is the night observed, cross-referenced
to Table 1. Fields with a 0 were not observed, and the intended field
centers are provided. The complete version of this table is in the electronic
edition of the Journal.  The printed edition contains only a sample.}
\end{deluxetable}

\begin{deluxetable}{ccccccc}
\tablewidth{0pt}
\tablecaption{{\VRI} Photometry \label{tab:data}}
\tablehead{
  \colhead{ID\tablenotemark{a}} &
  \colhead{$V$\tablenotemark{b}} &
  \colhead{$R_C$\tablenotemark{b}} &
  \colhead{$I_C$\tablenotemark{b}} &
  \colhead{R.A.} &
  \colhead{Decl.} &
  \colhead{Field(s)}
}
\startdata
lOri J051750.9+103824 & 16.738 & 15.993 & 16.838 & 05 17 50.916 & 10 38 23.84 & 146 \\
lOri J051751.2+102548 & 18.714 & 17.319 & 16.622 & 05 17 51.197 & 10 25 47.72 & 146 \\
lOri J051751.5+105655 & 15.396 & 14.862 & 14.282 & 05 17 51.502 & 10 56 55.41 & 146 \\
lOri J051751.5+103716 & 16.674 & 15.930 & 15.304 & 05 17 51.515 & 10 37 16.45 & 146 \\
lOri J051751.6+105902 & 17.676 & 17.070 & 16.314 & 05 17 51.593 & 10 59 01.69 & 146 \\
\enddata
\tablecomments{The complete version of this table is in the electronic
edition of the Journal.  The printed edition contains only a sample.}
\tablenotetext{a}{The identifier is constructed from the USNO-A2.0
coordinates.  For the 110 stars where this identifier is not unique,
we have appended an ``a'' for the westmost or a ``b'' for the eastmost
star to distinguish between them.}
\tablenotetext{b}{The mean photometric error is 0.032 ($12<R<16$), 0.069
($16<R<17$), or 0.12 ($17<R<18$) mag as described in the text.}
\end{deluxetable}

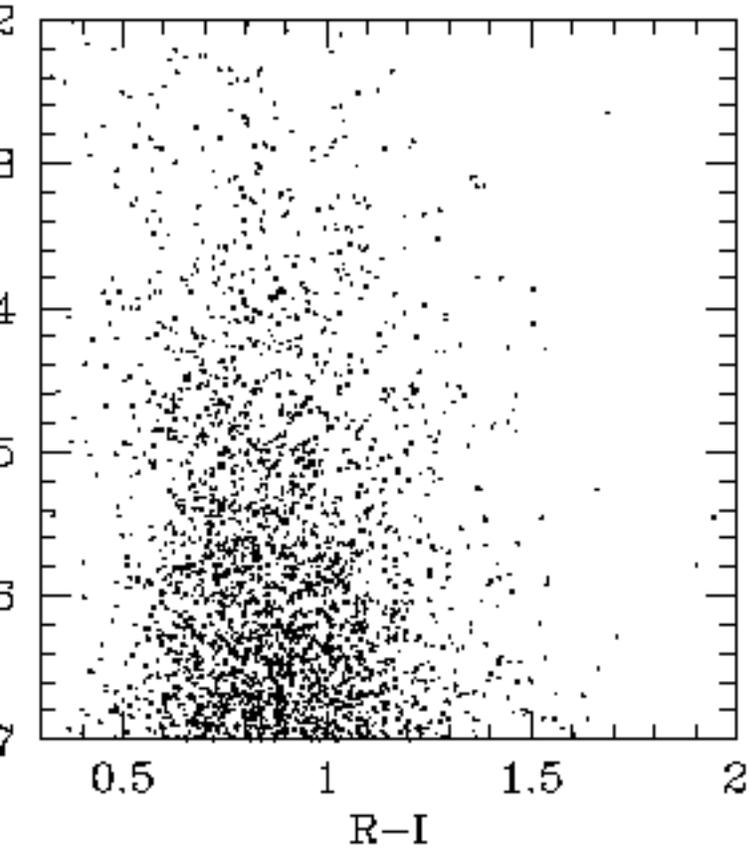
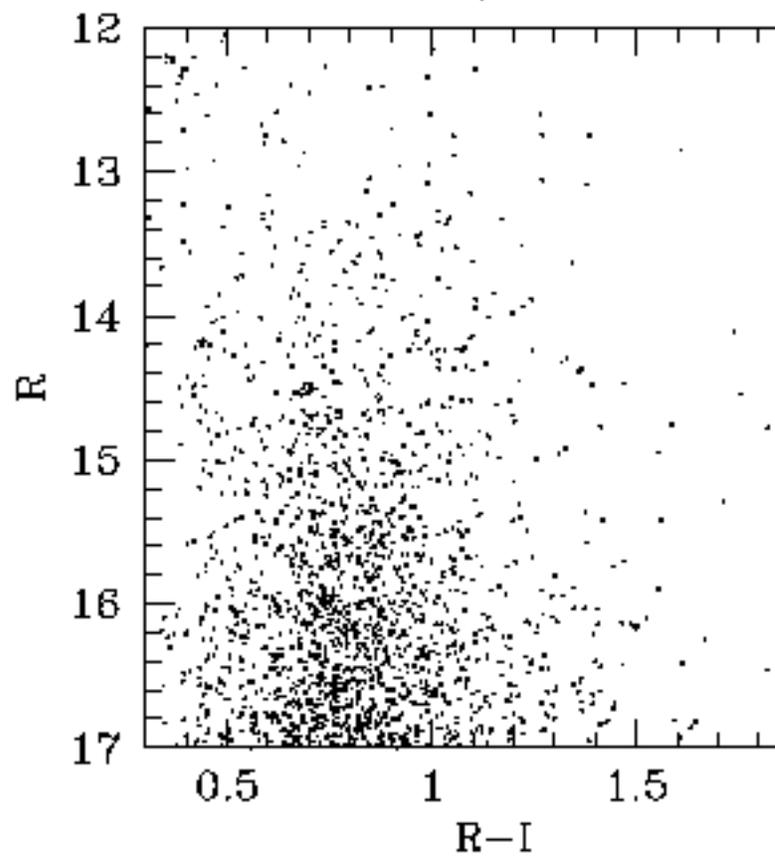
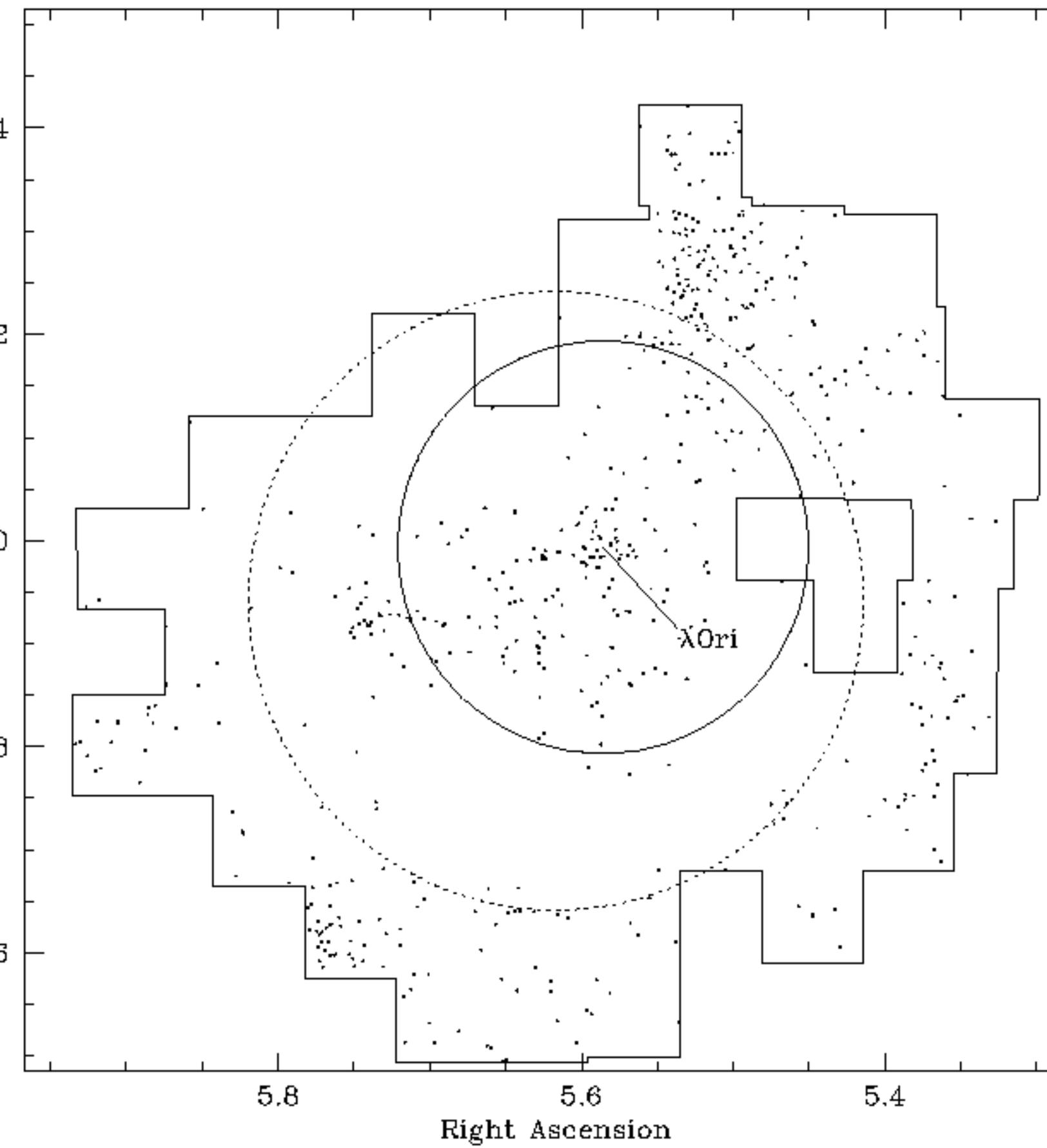
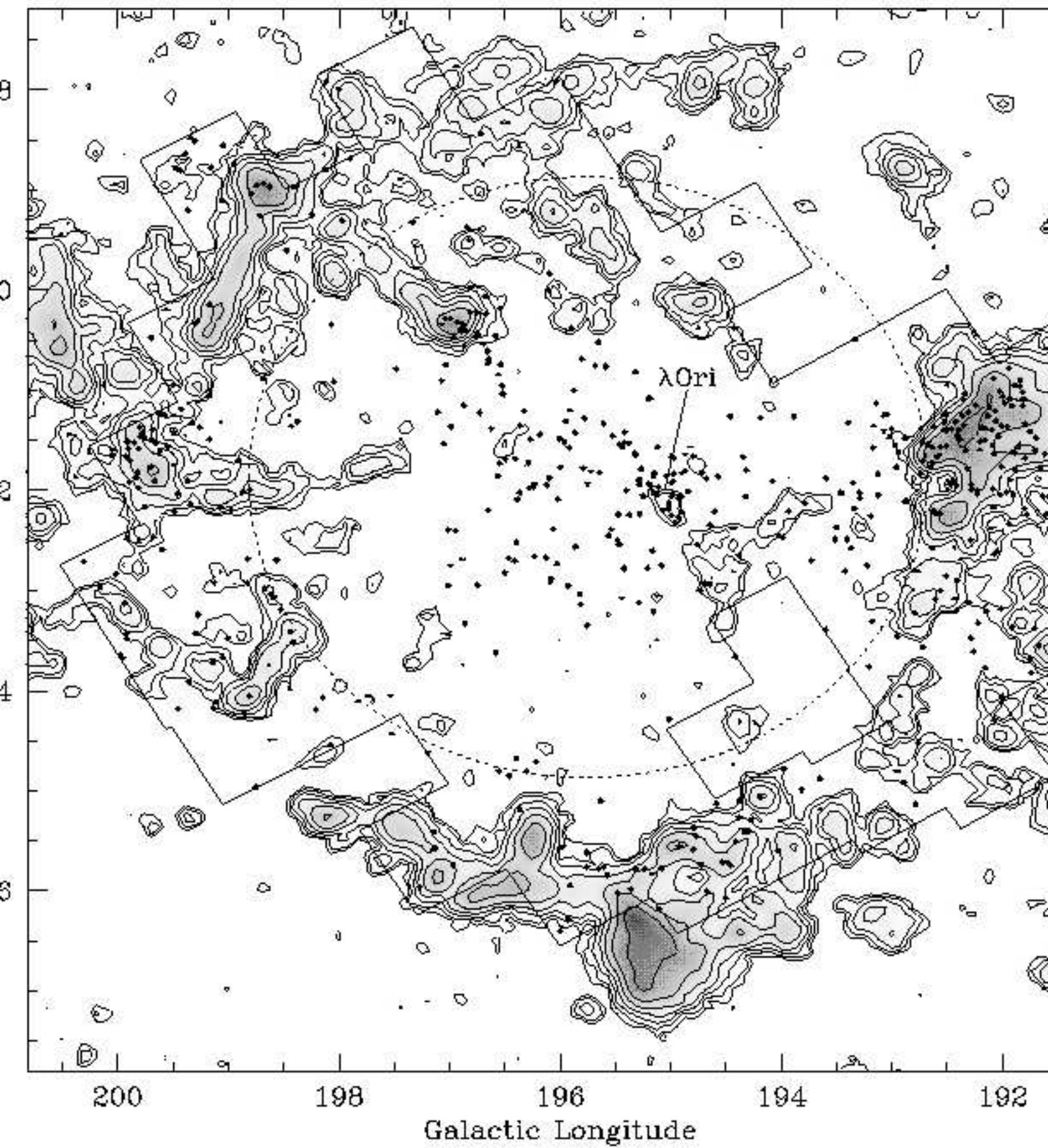
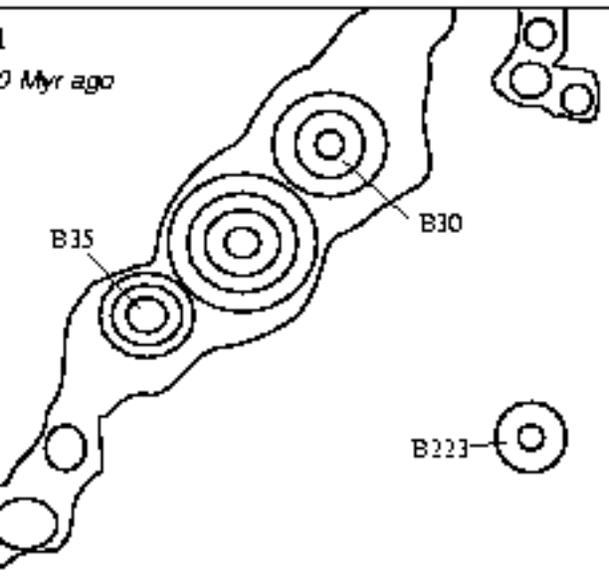
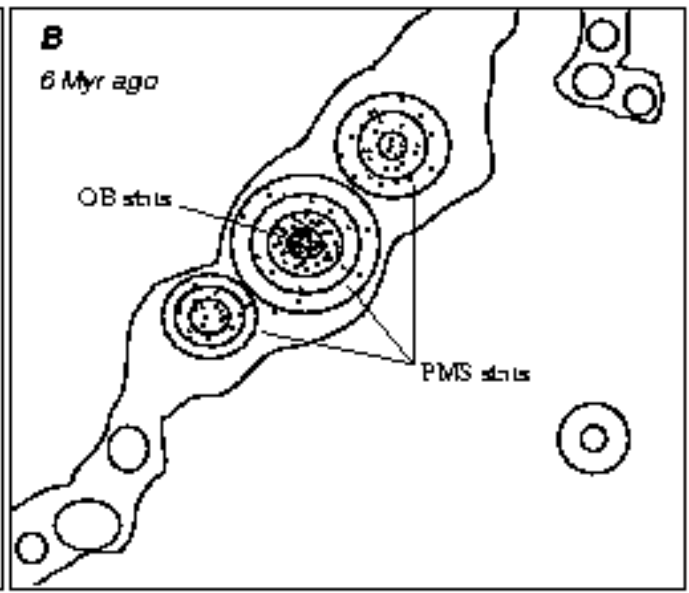
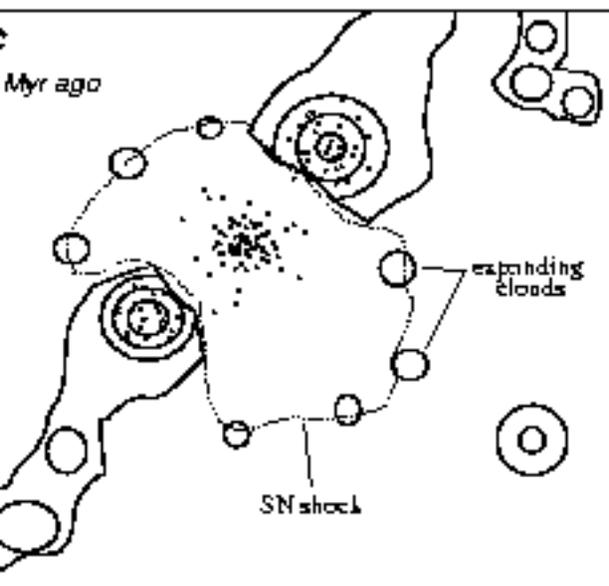
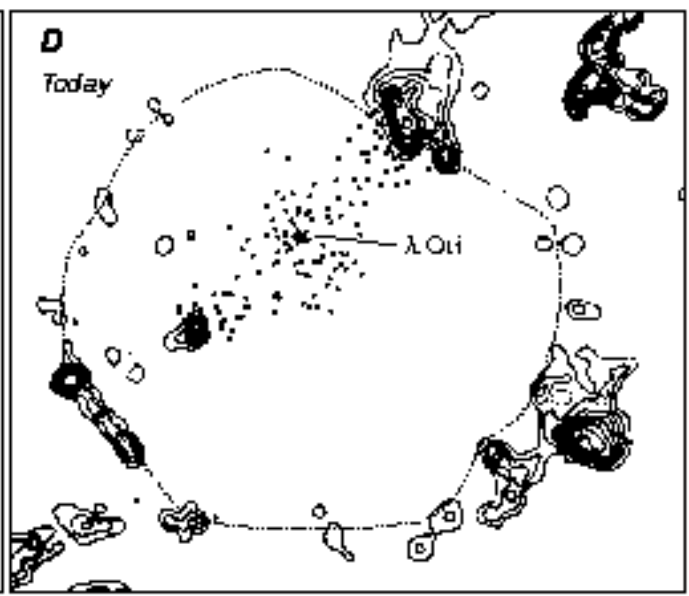
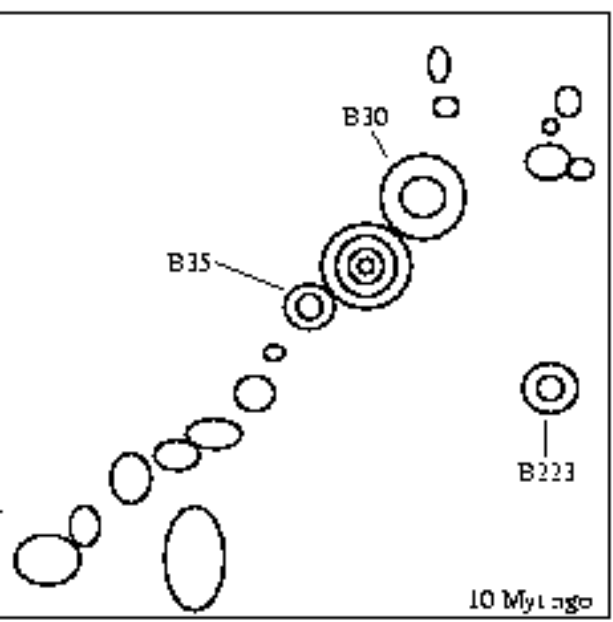
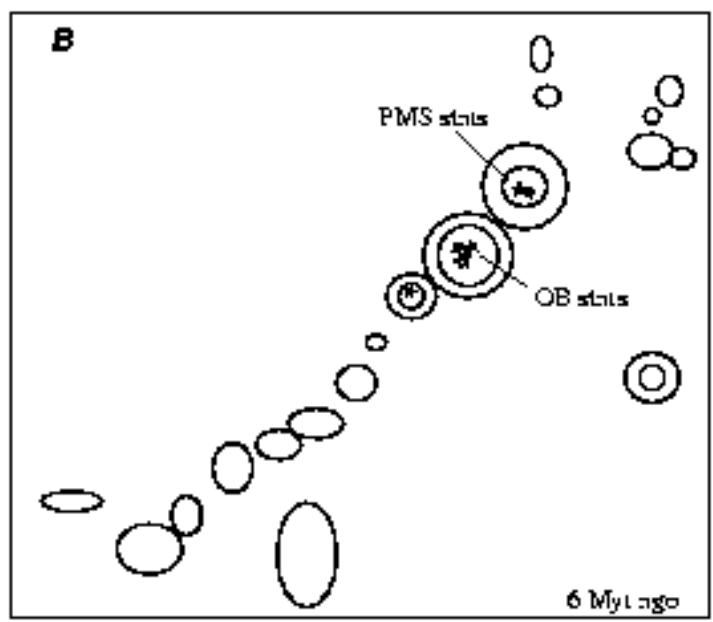
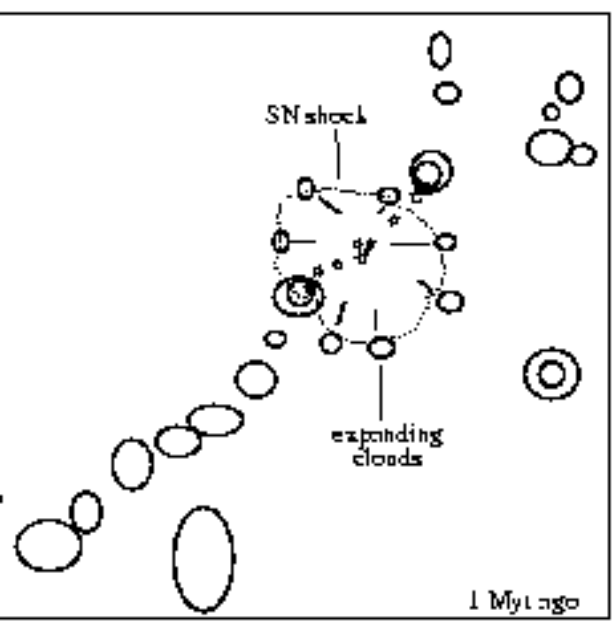
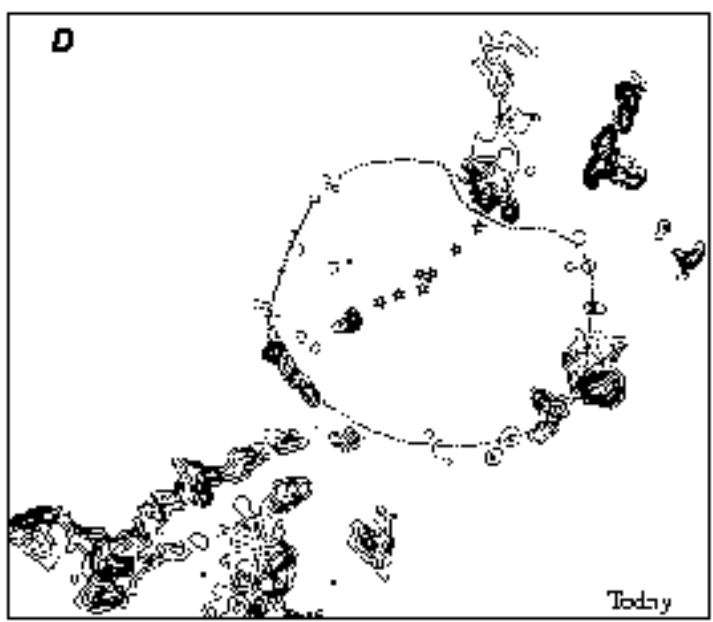
\end{document}